\documentstyle[12pt]{article}
\def\beq{\begin{equation}}
\def\eeq{\end{equation}}
\def\bea{\begin{eqnarray}}
\def\eea{\end{eqnarray}}

\def\ba{\begin{array}}
\def\ea{\end{array}}
\def\bce{\begin{center}}
\def\ece{\end{center}}

\def\nonu{\nonumber}

 \def\De{\Delta}

\def\vol#1{{\bf #1}}
\def\nupha#1{Nucl. Phys. \vol{#1} }
 
\def\phlta#1{Phys. Lett. \vol{#1} }

\def\PRL#1{Phys. Rev. Lett. \vol{#1} }

\begin{document}
\begin{titlepage}
\rightline{SNUTP-98-085, APCTP-98-019, UM-TG-210, BROWN-HET-1144}
\rightline{hep-th/9808143}
\def\today{\ifcase\month\or
January\or February\or March\or April\or May\or June\or
ppJuly\or August\or September\or October\or November\or December\fi,
\number\year}
%\rightline{\today}
\vskip 1cm
\centerline{\Large \bf The Large $N$ Limit of ${\cal N}=1$ Field Theories 
from F Theory}
\centerline{\Large \bf }
\vskip 1cm
\centerline{\sc Changhyun Ahn$^{a}$, 
Kyungho Oh$^{b}$ and Radu Tatar$^{c}$}
\vskip 1cm
\centerline{{\it $^a$ Rm NS1-126, Dept. of Physics, 
Kyungpook National University,
Taegu 702-701, Korea}}
\centerline{{\tt ahn@kyungpook.ac.kr}}
\centerline{{ \it $^b$ Dept. of Mathematics, University of 
Missouri-St. Louis,
St. Louis, MO 63121, USA }}
\centerline{{\tt oh@arch.umsl.edu}}
\centerline{{\it $^c$ Dept. of Physics, University of Miami,
Coral Gables, FL 33146, USA and} }
\centerline{{\it $^d$ Dept. of Physics, Brown University,
Providence, RI 02912, USA}}
\centerline{{\tt tatar@het.brown.edu}}
\vskip 1cm
\centerline{\sc Abstract}
\vskip 0.2in
We study AdS/CFT correspondence in four dimensional 
${\cal N}=1$ field theories realized on the worldvolume of
D3 branes near the intersections between  D7 and D7' branes in F theory
studied by
Aharony et al. We consider the compactification of F theory on 
elliptically fibered Calabi-Yau threefolds corresponding to two sets of 
parallel D7 branes sharing six spacetime directions. This can be viewed as
orbifolds of
six torus $T^6$ by ${\bf Z}_p \times {\bf Z}_q (p, q=2, 3, 4, 6)$. We find the 
large $N$ spectrum of chiral primary operators by exploiting the property of
AdS/CFT correspondence. Moreover, we discuss supergravity solutions for 
D3 branes in D7 and D7' branes background.    
\vskip 2.1in
\leftline{revised, Jan., 1999}
\end{titlepage}
\newpage
\setcounter{equation}{0}
%\setcounter{page}{0}

%****************************************************
%*****************************************************
%*****************************************************
\section{Introduction}
\setcounter{equation}{0}

In \cite{mal}, 
the large $N$ limit of superconformal field theories (SCFT) 
was described by taking the
supergravity limit for compactifications on anti-de Sitter (AdS) space.
The correlation functions of SCFT can be 
obtained from 
the supergravity action dependence on the fields at the
boundary of the AdS space \cite{fk,polyakov,witten}.
In particular, 
the ${\cal N}=4$ super $SU(N)$ Yang-Mills theory in 4 dimensions is
obtained in the compactification of
type IIB string theory on $AdS_5 \times {\bf S}^5$. The gauge group can be
replaced by $SO(N)/Sp(N)$ \cite{witten1} by taking
appropriate orientifold operations (See related works \cite{aoy,kaku,fs}).

There are ${\cal N} =2, 1, 0$ superconformal theories in 
4 dimensions
for which the corresponding supergravity description 
on orbifolds of $AdS_5 \times {\bf S}^5$
has been initiated in \cite{kachru,vafa}. This proposed duality was tested
by studying
the Kaluza-Klein (KK) states of supergravity theory and 
by comparing them with the chiral primary operators
of the SCFT on the boundary \cite{ot}.
Furthermore, the field theory/M theory duality gives rise after
compactification of
on $AdS_4$ or $AdS_7$ to some superconformal theories
in 3 and 6 dimensions, respectively. The maximally supersymmetric theories
have been studied in \cite{aoy,lr,minwalla2,halyo1,gomis} and
the lower supersymmetric case was also realized on the worldvolume of 
M theory
at orbifold singularities \cite{fkpz} (See also 
\cite{berkooz}).
Along the line of \cite{ot}, 
the KK states of supergravity theory on the orbifolds of
$AdS_4 \times {\bf S}^7$ were studied and compared with the chiral 
primary operators
of the SCFT on the boundary \cite{eg,aot1}. The analysis for 
orbifolds of $AdS_7 \times {\bf S}^4$ was worked out in \cite{aot}.
The KK spectrum description on the twisted states of $AdS_5$ orbifolds
was discussed in \cite{gukov}. 

Recently, it has been found \cite{afm} 
that in ${\cal N}=2, 1$ field theories by worldvolume of
D3 branes near D7 branes, their large $N$ spectrum of chiral primary operators
is computed by using their string theory duals. (Other related
papers are
\cite{gukov,kw,gubser,kakush,gk}).
In this paper, we discuss the following two things.
$i)$ First, we will compute the large $N$ spectrum of ${\cal N}=1$ SCFT in 
section 3 which
was not done in \cite{afm} explicitly.
$ii)$ Second, we discuss the qualitative property of supergravity solution
in the nonconformal case in section 4.  
 
In section 2, we review some known results for the moduli space of F theory
on elliptically fibered Calabi-Yau threefolds  where the coupling 
remains
constant. 
%We also write down the enhanced gauge symmetries arising from
%the intersection of two types of singularities.
In section 3, 
we can read the value for $k$ characteristic for the  irreducible 
representations of $SU(4)$ isometry group from the KK spectrum of 
original $AdS_5 \times {\bf S}^5$ from the metric  describing the D7 and
D7' brane background.
 The value for $k$ are determined by the three momenta
corresponding to the three angular variables coming from three ${\bf
S}^1$'s and 
two additional nonnegative integers \cite{afm}.
We can read off $U(1)_R$ charge  from the relation 
between
the dimension of chiral operators and $U(1)_R$ charge in the 
${\cal N}=1$ superconformal algebra. We see the
representations giving rise to chiral primary operators in the
boundary ${\cal N}=1$ CFT. 
Finally in section 4, we consider the solution of supergravity in nonconformal
case in D7 and D7' branes background qualitatively.

%*****************************************************
%*****************************************************
%*****************************************************
\section{F Theory on Calabi-Yau Threefolds at Constant Coupling }
\setcounter{equation}{0}
In this section we review the part of \cite{an} 
which is necessary for our present aim.
Let us consider the compactification of F theory on elliptically fibered 
CY 3-fold over ${\bf CP}^{1} \times {\bf CP}^{1} $
where the coupling is
constant over the base. 

Let $z$ and $w$ be  the affine coordinates~\footnote{
We define  
$z\equiv x^8+ix^9$ and $ w \equiv
x^6+ix^7$ in section 3.}
 of the base 
${\bf CP}^{1} \times {\bf CP}^{1} $. Then the elliptically fibered 
CY 3-fold  can be described \cite{MV,GJ2} in the Weirstrass form 
$$
y^2  = x^3 + f(z,w) x +g(z,w)
$$
where   $f$ and $g$ are the 
polynomials of degree $8$, $12$ respectively in each of the variables.
The $j$-invariant of   the modular parameter $\tau(z,w)$ of the fiber
is  given by
\beq
j(\tau(z,w))=\frac{4(24 f(z,w))^3}{\De(z,w)},
\eeq
where the discriminant $\De(z,w)$ is 
$$
\De(z,w)=4(f(z,w))^3+27(g(z,w))^2.
$$
The fibers are degenerated when $\De(z,w) = 0$ and 
the type IIB 
7-branes  are located at these points. The $j$-invariant is
invariant under $SL(2, {\bf{Z}})$ transformation of $\tau$ and
determines the complex structure of the elliptic fiber.

By requiring the modular parameter $\tau(z,w)$ to be constant up to an
$SL(2, {\bf{Z}})$ transformation, we obtain $f^3 \sim g^2$.
To simplify the problem, we will assume that $f$ and $g$
are factorized. That is, 
$f(z,w)=\alpha f_1(z) f_2(w)$ and $g(z,w)=g_1(z) g_2(w)$ where $\alpha$
is a constant. Then we have the following possible cases.

$ \mbox{Case} \; 1): \;\;\;  {\bf T}^6/{\bf Z}_2 \times {\bf Z}_2$

The solution for
constant modulus obtained by rescaling $y$ and $x$
and setting the overall coefficient to be 1 has been found in
\cite{Aharony} in the form:
\bea
&&f_1(z)=\prod_{i=1}^{4} (z-z_i)^2, \;\;\; f_2(w)=\prod_{i=1}^{4} (w-w_i)^2, 
\nonu \\
&&g_1(z)=\prod_{i=1}^{4} (z-z_i)^3, \;\;\; g_2(w)=\prod_{i=1}^{4} (w-w_i)^3,
\eea
where $z_i$'s and $w_i$'s are constants.
%This special compactification corresponds to a configuration where
%the 24 D7 branes are grouped into 4 sets of 6 coincident D7 branes located at
%the points, $z_1, z_2, z_3, z_4$. Around each of fixed points 
%$z_i \in \bf{CP}^1$, there exists an $SL(2,{\bf Z})$ monodromy
%$$
%\left( \begin{array}{rr}
%       -1& 0\\
%        0& -1
%\end{array}
%\right)
%$$
%acting on the homology $H_1({\bf T}^2 \times {\bf S}^2, \bf{Z})$ of the fiber
%over ${\bf CP}^1$ corresponding to the hyperelliptic
%involution of the elliptic part  ${\bf T}^2$ of the fiber. 
%The same is true at the points $w=w_i$.
%We have $D_4$ singularities on the fiber located at
% $z=z_i$ or  $w=w_i$ which gives
%rise to an enhanced $SO(8)$ gauge symmetry.
%The spacetime theory is an ${\cal N}=1$ supersymmetric theory whose field
%content was found in \cite{BZ,GM}. For example,
%the open string sectors lead to $SO(8)$ gauge group for each D7 branes 
%coming 
%from
%two ${\bf Z}_2$ factors for a total enhanced gauge symmetries 
%$(SO(8))^8$ \cite{GM}.
 
$\mbox{Case} \; 2): \;\;\; {\bf T}^6/{\bf Z}_3 \times {\bf Z}_2$

As pointed out in \cite{DM}, in the limit of
$\alpha \rightarrow 0$, we get 
$j(\tau(z,w))=0$ from which $\tau(z,w)=e^{\frac{\pi i}{3}}$.
The polynomials are given by
\bea
&& f_1(z)=0, \;\;\; g_1(z)=\prod_{i=1}^3 (z-z_i)^4, \nonu \\
&& f_2(w)=\prod_{i=1}^4 (w-w_i)^2, \;\;\; g_2(w)=\prod_{i=1}^4 (w-w_i)^3,
\label{eq:z3}
\eea
where the 12(12) zeroes of $g_1(z)(g_2(w))$
coalesce into 3(4) identical ones of order 4(2) each.
%In this case, the discriminant, $\Delta(z,w)$ takes the form of
%$
%\Delta(z,w)=27 \prod_{i=1}^3 (z-z_i)^8 \prod_{j=1}^4 (w-w_j)^6.
%$
%The singularity type from 
%Tate's algorithm \cite{Tate}
%at a zero of the discriminant gives rise to the enhancement of gauge 
%symmetries \cite{MV2}.
%Each point $z=z_i$ on the first ${\bf CP}^1$ 
%factor carries a deficit angle of 
%$3 \pi/2$ all three of them together deforming the  ${\bf CP}^1$ to
%${\bf T}^2/{\bf Z}_3$. For each point $w=w_i$ on the second ${\bf CP}^1$, 
%there
%is a deficit angle of $\pi$ all four of them deforming ${\bf CP}^1$ to 
%${\bf T}^2/{
%\bf Z}_2$. This is related to orientifold of F-theory on 
%${\bf T}^6/{\bf Z}_3 
%\times {\bf Z}_2$.
%The full enhanced gauge symmetry group is $(F_4)^3 \times 
%(G_2)^4$ by resolving the singularity for each point of $z=z_i$ and $w=w_i$.

$\mbox{Case} \; 3): \;\;\; {\bf T}^6/{\bf Z}_3 \times {\bf Z}_3$

When the 12 zeroes of $g_2(w)$ have coalesced into 3 identical ones of
order 4
and the ones of $g_1(z)$ are given as in eq.(\ref{eq:z3}),
we have the following:
\beq
 f_1=0, \;\;f_2=0, \;\; g_1(z)=\prod_{i=1}^3 (z-z_i)^4, \;\; g_2(w)=
\prod_{i=1}^3 (w-w_i)^4,
\label{eq:33}
\eeq
where the discriminant is given by
$
\Delta(z,w)=27 \prod_{i=1}^3 (z-z_i)^8 \prod_{j=1}^3 (w-w_j)^8.
$
%Each point $w=w_i$ on the second ${\bf CP}^1$ factor has a 
%deficit angle of 
%$3 \pi/2$ all three of them together deforming the  ${\bf CP}^1$ to
%${\bf T}^2/{\bf Z}_3$. 
%The total enhancement of  gauge symmetry group is given by
%$ (E_6)^6 \times (SU(3))^3$. 

$ \mbox{Case} \; 4): \;\;\; {\bf T}^6/{\bf Z}_3 \times {\bf Z}_6$

If the 12 zeroes of $g_2(w)$ has coalesced into 3 zeroes of order 5, 4, 3
each and those of $g_1(z)$ are the same as before like (\ref{eq:33}),  
it is easy to see that
we have the following: 
\beq
f_1=0, \;\; f_2=0, \;\; g_1(z)=\prod_{i=1}^3 (z-z_i)^4, \;\; g_2(w)=
(w-w_1)^5(w-w_2)^4(w-w_3)^3.
\eeq
Each point $w=w_i$ on the second ${\bf CP}^1$ factor has a deficit angle of 
$5 \pi/3, 4 \pi/3$ and $\pi$ all  
together deforming the  ${\bf CP}^1$ to
${\bf T}^2/{\bf Z}_6$. 
%The gauge symmetry group is $(F_4)^3 \times (G_2)^2 \times 
%SU(2) \times SU(3)
%\times E_8 \times E_6 $ gauge symmetry appears.

$\mbox{Case} \; 5): \;\;\; {\bf T}^6/{\bf Z}_4 \times {\bf Z}_2$

Another possibility is as follows:
\bea
&& f_1(z)=(z-z_1)^3(z-z_2)^3(z-z_3)^2, \;\;\; g_1(z)=0, \nonu \\
&& f_2(w)=\prod_{i=1}^ 4 (w-w_i)^2, \;\;\;
g_2(w)=\prod_{i=1}^4 (w-w_i)^3,
\label{eq:z4}
\eea
which corresponds to $\tau=i$ from $j(\tau(z,w))=13824$.
%This time it can be easily checked that the discriminant is given by
%$
%\Delta(z,w)=4 (z-z_1)^9(z-z_2)^9(z-z_3)^6 \prod_{i=1}^ 4 (w-w_i)^6.
%$
%The singular fiber over each fixed points $z_1, z_2$ is of $E_7$ type. 
%The other singular
%fiber over $z_3$ is of $SO(8)$ type.
%At the intersection points near $z=z_1$ and $w=w_1$, the gauge group
%appears to be $E_7 \times SU(2) \times SO(8)$ enhanced by an extra 
%$SU(2)$ factor.
%$(E_7)^2  \times SO(7) \times (SU(2))^2 \times (SO(8))^4$ gauge 
%symmetry appears there.

$\mbox{Case} \;6): \;\;\; {\bf T}^6/{\bf Z}_4 \times {\bf Z}_4$

Finally, we have the case when the 8 zeroes of $f_1(z)(f_2(w))$ coalesce 
into 3 of orders
3, 3 and 2 respectively and $g_1=g_2=0$
\bea
&& f_1(z)=(z-z_1)^3(z-z_2)^3(z-z_3)^2, \;\;\;
g_1(z)=0, \nonu \\
&& f_2(w)=(w-w_1)^3(w-w_2)^3(w-w_3)^2, \;\;\; g_2(w)=0.
\eea
%In this case, the gauge group appears $(E_7 \times SU(2))^2 \times SO(7)$ 
%at the intersections of two $E_7$ singularities.
%Then we get
%the enhanced gauge symmetry group 
%$(E_7)^4  \times (SO(7))^6 \times(SU(2))^8$. 

Note that there are other two  cases,
the intersection of $SO(8)$ and $E_8$, the intersection of
$E_8$ and $E_8$ which are ruled
out because they violate the CY 3-fold condition. The intersection of 
$E_6$ and $E_7$ and the intersection of $E_7$ and $E_8$ do not live in
F theory moduli space where the couplings remain constant.
In the next section, we will consider how the above six cases can be
realized
in the context of AdS/CFT correspondence and how to obtain predictions for
the full
large $N$ spectrum of chiral primaries.

%*****************************************************
%*****************************************************
%*****************************************************
\section{The Surviving Kaluza-Klein Spectrum}
\setcounter{equation}{0}

Let us take $N$ D3 branes with the worldvolume along $(x^0,
x^1, x^2, x^3)$, the appropriate number of D7 branes
with worldvolumes along $(x^0, x^1,
x^2, x^3, x^4, x^5, x^6, x^7)$, and D7' branes in $(x^0, x^1, x^2, x^3,
x^4, x^5, x^8, x^9)$. 
By adding D3 branes into D7 and D7' brane system,the worldvolume 
field theory has ${\cal N}=1$ supersymmetry.
The position of the D7 brane is given by $z$, while the
D7' brane position is given by $w$. As usual, O7 plane and O7' plane
have their worldvolume parallel with the ones of the D7 and D7' brane
respectively. 
\bea
D3 & :  & (x^0,x^1, x^2, x^3)  \nonu \\
D7/O7 & :  & (x^0, x^1,x^2, x^3, x^4, x^5, x^6, x^7) \nonu \\ 
D7'/O7' & : & (x^0, x^1, x^2, x^3, x^4, x^5, x^8, x^9).
\eea

Let us study the F theory on the vicinity of a
 CY 3-fold singularity
(local F theory geometry) which corresponds to 
coincident D7 and D7' branes in type IIB theory where the complexified
coupling (modular parameter of the fiber) $\tau(z, w)$ 
is a constant. 
The metric describing the D7 and D7' brane background looks like
$AdS_5 \times {\bf S}^5$ and can   
be (up to a constant normalization) summarized by: 
\bea 
\mbox{$SO(8) \times SO(8)$:}\quad ds^2 & = &\mid z^{-\frac{1}{2}}dz\mid^2+
 \mid w^{-\frac{1}{2}}dw\mid^2,\nonumber \\ 
\mbox{$E_6 \times SO(8)$:}\quad ds^2 & = &\mid z^{-\frac{2}{3}}dz\mid^2+
\mid w^{-\frac{1}{2}}dw\mid^2
 ,\nonumber \\ 
\mbox{$E_6 \times E_6$:}\quad ds^2 & = &\mid z^{-\frac{2}{3}}dz\mid^2+
\mid w^{-\frac{2}{3}}dw\mid^2
 ,\nonumber \\ 
\mbox{$E_6 \times E_8$:}\quad ds^2 & = &\mid z^{-\frac{2}{3}}dz\mid^2+
\mid w^{-\frac{5}{6}}dw\mid^2
 ,\nonumber \\ 
\mbox{$E_7 \times SO(8)$:}\quad ds^2 & = &\mid z^{-\frac{3}{4}}dz\mid^2+
 \mid w^{-\frac{1}{2}}dw\mid^2
 ,\nonumber \\ 
\mbox{$E_7 \times E_7$:}\quad ds^2 & = &\mid z^{-\frac{3}{4}}dz\mid^2+
 \mid w^{-\frac{3}{4}}dw\mid^2.
 \label{metric}
\eea 
The metric can be explained in the same way as in \cite{fs}. 
These spaces 
are described as the orbifolds ${\bf C}^2/{\bf Z}_p \times {\bf Z}_q$ 
with $p, q=2,3,4,6$ that have been already considered in section 2.  Thus  
the covering space is the complex $u, v$ surface, with $z =
u^p$ and $w=v^q$. Now by descending the Euclidean metric 
$ds^2 = \mid du\mid^2 + 
\mid dv\mid^2 $ on the covering space, we obtain the desired metric 
which is equivariant under the orientifold action
 $u\longrightarrow \exp{(2\pi i/p)} u$
and $v\longrightarrow \exp{(2\pi i/q)} v$.

 At various types of 
intersection points there exists an enhancement of gauge symmetries which
gives rise to conformal field theories on the D3 branes.
For example when one considers Case 1) 
for which there are $SO(8)$ gauge fields living both on the D7 and
D7' branes, there exists an $SO(8) \times SO(8)$ global symmetry on the 
D3 branes. 

We can choose coordinates so that the angular part of (\ref{metric}) 
becomes \cite{afm}
\bea
ds^2 = d\theta^2 
+ \cos^2\theta d\phi_1^2 + \sin^2\theta \left[d\psi^2 +
\cos^2\psi d\phi_2^2 +  \sin^2\psi d\phi_3^2 \right],
\eea
with $\alpha$ deformed boundary conditions which will appear later and 
monodromies
in the $\phi_1$ and $\phi_2$
variables. We have various types of singularities at $\sin\theta=0$
which is an ${\bf S}^3$ in the compact space. This means two intersection
of singularity types coming from
D7 and D7' branes along two ${ \bf S}^3$ in 
the compact space which intersect along an ${\bf S}^1$.

The Laplacian in this metric is
\bea
\nabla^2 & = & { 1\over \sin^3\theta\cos\theta} { d \over
d \theta} \sin^3\theta\cos\theta {d \over
d \theta }+
{1\over \sin^2\theta} { d \over d\psi} \sin\psi\cos\psi {d \over
d\psi} \nonu \\
  & & 
 - {m_1^2 \over \cos^2\theta } - { m_2^2 \over \sin^2\theta\cos^2 \psi }
-{m_3^2 \over \sin^2\theta\sin^2\psi},
\eea
where $m_1^2, m_2^2$ and $m_3^2$ are the eigenvalues of the Laplacian on 
three ${\bf S}^1$'s 
respectively. The $SU(4)$ isometry corresponding to the global
symmetry in 4 dimensional field theory breaks into $ U(1) \times
U(1) \times U(1)$ due to the presence of D7 and D7' branes.
The $U(1)_R$ charge in the ${\cal N}=1$ superconformal algebra 
can be written as the sum of the charges for the $SO(2)$'s  
acting on $\phi_1,\phi_2,\phi_3$,
so it is:
\bea
U(1)_R=2 (m_1 + m_2 + m_3).
\label{R}
\eea
The spherical harmonics, denoted by $Y(m_1, m_2, m_3, m, n)$ 
are classified by the momenta
$m_1,m_2$ and $m_3$ in the $\phi$ variables and by two additional
non-negative 
integers $m$ and $n$, such that the eigenvalue of the total 
Laplacian \cite{afm} 
is
$k(k+4)$ where $k = |m_1| + |m_2| + |m_3| + 2m + 2n$. 
In the ${\bf S}^5$ case all these numbers are integers. 
In the present case 
the periodicity conditions on $\phi_1,\phi_2$ are different,
so $m_i = {\widetilde m_i}/(1-\alpha_i/2)$ ($i =1,2$), 
but $m_3,n$ and $m$ are still integers. 
The values of $\alpha_i$'s are corresponding to each 
singularity types: 1, 4/3, 3/2, 5/3 
for
$SO(8), E_6, E_7$ and $E_8$. Note that
only six combinations are possible as discussed in section 2.
We have the eigenfunctions $ e^{i 
{\widetilde m_i}\phi/(1-\alpha_i/2)}$ rather than $e^{i m \phi} $. Then
\bea
k = {|{\widetilde m_1}|\over(1-\alpha_1/2)} +
{|{\widetilde m_2}|\over(1-\alpha_2/2)} +
 |m_3| + 2m+ 2n.
\eea
From the allowed values for $\alpha_{i}$, we see that $k$ has always
integer values.

We are now ready to compare the KK spectrum of supergravity and
corresponding chiral
primary operators living on the boundary of $AdS_5$. Using the
relation between the dimensionsof the $R$-symmetry representations, we
can
see that the condition to obtain a chiral primary field in the
superconformal algebra is $m = n = 0$. In this case, 
\bea
k = {|{\widetilde m_1}|\over(1-\alpha_1/2)} +
{|{\widetilde m_2}|\over(1-\alpha_2/2)} +
 |m_3|
\eea
We can now make a discussion over some surviving KK states.
This can be
done
by decomposing $SU(4)$ representations of KK states of  four dimensional
${\cal N}=4$ supersymmetry under representations of the smaller global
symmetry group of our theory with ${\cal N}=1$ and keeping only states
with right $R$ symmetry charges which are determined by scaling dimensions of
corresponding chiral primary operators.

$\bullet$ Scalar fields with Dynkin labels $(0, k, 0), \;\;\;
k=2, 3, 4, \cdots$
 
Let us consider the case of $k=2$ which would correspond to one non-zero
$\widetilde{m}$ or to $m_3 = 2$. 
How does the representation ${\bf 20'}$ of 
$SU(4)$ ?  The corresponding dimension $\Delta=k=2$ of the chiral primary 
operator has $R$ charge $4/3$. Obviously there is no solution for nonnegative
integers ${\widetilde m_1}, {\widetilde m_2}$ and $m_3$ satisfying this
constraint. There is no any dimension $2$ chiral 
primary operators in the boundary ${\cal N}=1$ CFT.
We can go on further to $k=3$.
How does the representation ${\bf 50}$ of 
$SU(4)$ go? The dimension  $\Delta=k=3$ chiral primary operator has $R$ 
charge $2$. One of the solutions, $m_1=m_2=1,
m_3=-1$ which will give the right number of $k$. Since 
it does not depend on the values of $\alpha_1$ and $\alpha_2$,
all the six possibilities we had in section 2 are valid in this case.
So we expect dimension $3$ chiral 
primary operators in the boundary ${\cal N}=1$ CFT. That is, there exists 
a state given by the spherical harmonics, $Y(1,1,-1,0,0)$. 
What happens for $k=4$ ? That is, ${\bf 105}$. 
The corresponding dimesion $\Delta=k=4$ chiral primary 
operator has $R$ charge $8/3$. There is no solution for 
${\widetilde m_1}, {\widetilde m_2}$ and $m_3$ satisfying this
constraint like as the previous consideration.

$\bullet$ Scalar fields with Dynkin labels $(0, k, 2), 
\;\;\;k=0, 1, 2, \cdots$

Take the representation ${\bf 45}$ of $SU(4)$ ? This is 
the case $k=1$.
The dimension $\Delta=k+3=4$ chiral primary operators of 
the boundary theory should have $U(1)_R$ charge $8/3$. Again this does not 
lead to any chiral primary operators in the boundary.

So far we have studied the chiral primaries of ${\cal N}=1$ SCFT from bulk
modes. 
In addition
to that, there also exists a contribution from fields 
living at the singularities.
The spectrum of operators is known when we consider the intersection of
two
$SO(8)$ singularities which is ${\bf Z}_2 \times {\bf Z}_2$  orientifold
of the type IIB string theory
\cite{Aharony}. For $N$ D3 brane probes
the theory on D3 branes is $Sp(2N) \times Sp(2N)$ gauge theory with
two chiral superfields in ${\bf (2N, 2N)}$ representation from the D3 brane
coordinates along the $(x^6, x^7, x^8, x^9)$ directions, superfields in the
$({\bf N(2N-1)-1, 1) \oplus (1, 1) \oplus (1, N(2N-1)-1) \oplus (1, 1)}$ from
the D3 brane coordinates in the $(x^4, x^5)$, 
eight ${\bf (2N, 1)}$ which arises
from D7 branes located at a particular value of $z$ and ${\bf (1, 2N)}$
from the ones at particular value of $w$, superfields from the D3-D7 strings.
However, it is not known how to write down appropriate superpotential
which preserve the gauge symmetry and the global symmetry. Unlike the 
case of single singularity \cite{afm}, it is not clear how to compare
the analysis of KK mode on the ${\bf S}^1$ with the corresponding field 
theory.

%*****************************************************
%*****************************************************
%*****************************************************

\section{Supergravity Solution for 3 Branes}
\setcounter{equation}{0}

So far we considered particular limit of orbifold approach where the metric is
known. In this section 
we will comment on some qualitative features of D3 brane solution
in the background of D7 and D7' branes. 
Let us assume that there exists a supergravity solution: 
\bea
ds^{2}= dx_{\parallel}^2 + {g}_{ij}dx^{i}dx^{j},
\eea
where $dx_{\parallel}^2$ is the flat Minkowski 
metric in the directions $(x^0, x^1, x^2, x^3)$.
As before, we realize this
geometrically as an elliptically fibered CY 3-fold over a base
${\bf CP}^1 \times {\bf CP}^1 $. 
The metric $g$ is a K\"ahler metric and the complexified
 IIB coupling $\tau(z, w) = \chi + ie^{-\phi}$ 
determines the modular invariants of the
elliptic fiber of CY 3-fold over the base ${\bf CP}^1 \times {\bf CP}^1 $. 
We assume that the metric of the base 
$ds^2_{Base}$ is diagonal   
\bea
ds^2_{Base} =g_{z {\bar z}}(z, \bar{z})dz d{\bar z} +
g_{w {\bar w}}(w, \bar{w})dw d{\bar w}.  
\eea
Here we denote the functional dependence of 
$g_{z {\bar z}}$ and $g_{w {\bar w}}$ explicitly. 
So far
a general solution for this metric is not known. However there
is a partial attempt to get a particular solution. For example,
by solving the Einstein equations,
Asano \cite{asano} provided  a
metric 
$
g_{z {\bar z}} =
\mbox{Im}\tau \mid \eta^2  
\prod_{i=1}(z-z_i)^{-1/12} \mid^2, \;\;\;
g_{w {\bar w}} =
 \mid \prod_{i=1}(w-w_i)^{-1/2} \mid^2, 
$
where the Dedekind $\eta$ function is given by $
\eta(\tau)=q^{1/24}\prod_{n=1}^\infty(1-q^n), \;
q=\exp(2\pi i \tau) $.  
 The positions of
D7 and D7' branes are given by $z_i$ and $w_i$ and the modular invariant
$\tau(z,w)$ is a function of $z$ only for this solution.
Since the fiberation itself is not symmetric w.r.t.  $z$ and $w$,
it is not strange to have an asymmetric metric of the base.
One can  see that both $ g_{z {\bar z}}dz d\bar{z}$ and $g_{w 
{\bar w}}dw d\bar{w}$
describe the metric on ${\bf CP}^1$ globally
because 
$g_{z {\bar z}}\rightarrow (z\bar{z})^{-2}$ as $z\rightarrow \infty$ and 
$g_{w {\bar w}}\rightarrow (w\bar{w})^{-2}$ as $w\rightarrow \infty$.

Let us introduce D3 branes into this
problem: this can be  done in such a way as to respect the identification
under the orientifold group alternatively. The solution for D3 branes 
reads \cite{keha} 
\bea
ds^2  =  f^{-1/2}dx_{\parallel}^2 + f^{1/2} {g}_{ij}dx^{i}dx^{j},  \;\;\;
%f  = 1 + \frac{4\pi gN\alpha '^2}{r^4}\nonumber
\eea
%where
%\bea
%dx_{\parallel}^2 = -(dx^0)^2 + \sum_{k=1}^{3} (dx^k)^2 , \;\;\;
%dx_\bot^2 = \mid du\mid^2 + \mid dv\mid^2  + (dx^4)^2 +(dx^5)^2
%\eea
and the five form field $F_{0123i}$ is given by 
$F_{0123i}=-\frac{1}{4} \partial_i f^{-1}$ and $f(x^i)$ is a function
of the coordinates transverse to D3 branes.
%The solution in the presence of the D7 brane-O7 plane system
%is obtained simply by identifying $u\rightarrow u\exp (2\pi i/p)$. 
%Similarly,
%the one for 
%the D7' brane-O7' plane system
%is obtained by identifying $v\rightarrow v\exp (2\pi i/q)$. 
Then we have to solve 
\bea
\frac{1}{\sqrt{g}} \partial_i ( \sqrt{g} g^{ij} \partial_j f )
\sim -{\sqrt{g}} N
\delta^6 (x-x^0),
\label{solve}
\eea
where the right hand side indicates a source term at the position of 
$N$ D3 branes.
The equation for this leads to
\bea
\left[ g_{z \bar{z}} g_{w \bar{w}} \nabla_y^2 +2 (g_{w \bar{w}} \partial_z 
\partial_{\bar{z}}+ g_{z \bar{z}}
\partial_w \partial_{\bar{w}} ) \right] f \sim - N \delta^2(z-z_0) 
\delta^2(w-w_0)
\delta^2(y).
\label{sol}
\eea
It is difficult to find an exact solution for this so we will solve it
perturbatively.
Let us consider the case of regular metric at $z_0$ and $w_0$
and in the new variables, $\widetilde{z}$ and $\widetilde{w}$ they satisfy
$\widetilde{z}_0=\widetilde{w}_0=0$. Then the solution for (\ref{sol})
can be obtained by expanding $g$ and $f$ like as
$g=1+c(|\widetilde{z}|^2+|\widetilde{w}|^2) +
\cdots$ and $f=f_0+f_1+ \cdots$, iteratively.
One can find 
\bea
f_0  \sim \frac{N}{(y^2+2|\widetilde{z}|^2 +2|\widetilde{w}|^2 )^2}
\eea
which behaves as $N/y^4$ for small $\widetilde{z}$ and $\widetilde{w}$
and
\bea
\left[\nabla_y^2+2 \partial_{\widetilde{z}} \partial_{\bar{\widetilde{z}}}+2
\partial_{\widetilde{w}} \partial_{\bar{\widetilde{w}}} \right] f_1=
-c \left( |\widetilde{z}|^2+|\widetilde{w}|^2 \right) \nabla_y^2 f_0.
\eea
From this the expression for $f_1$ is given by 
\bea
f_1 \sim -\frac{4c}{3} \frac{|\widetilde{z}|^4+|\widetilde{w}|^4 +
2|\widetilde{z}|^2 |\widetilde{w}|^2}{(y^2+2|\widetilde{z}|^2+
2|\widetilde{w}|^2)} f_0
\eea
which is small when both $\widetilde{z}$ and $\widetilde{w}$ are near
zero value.
Plugging these values back all other terms in $f$ can be obtained.

One can also study D3 branes moving on an ALE space or near conifold singularity.
A theory away from the conformal point can be obtained 
by moving D3 branes away. Again we need to solve (\ref{solve}) in this background.
In principle we can generalize our present method to three types of parallel D7 branes
with any two sharing 6 dimensions  and all of them sharing 4 dimensions
keeping  4 supercharges (this can be realized by F theory on Calabi-Yau fourfold).
In this case the supersymmetry can be preserved when we add
extra D3 branes in appropriate directions. It would be interesting to
elaborate this further.

%*****************************************************
%****************************************************
%*****************************************************
\vspace{2cm}

\centerline{\Large \bf Acknowledgments} 

We would like to thank J. Maldacena for very important comments in a 
previous version of this paper.
We thank A. Fayyazuddin for helpful e-mail correspondence.
This work of CA 
was supported (in part) by the Korea Science and Engineering 
Foundation (KOSEF) through the Center for Theoretical Physics (CTP) at Seoul 
National University. CA thanks Asia Pacific Center for Theoretical 
Physics for partial support and hospitality 
where this work has been started. 

%*****************************************************
%*****************************************************
%*****************************************************

\end{document}